\documentclass[aps,prb,floatfix,twocolumn]{revtex4} 
\usepackage{graphicx} \usepackage{amsmath} \usepackage{amssymb}

\newcommand{\BEQ}{\begin{eqnarray}} 
\newcommand{\EEQ}{\end{eqnarray}} 
\newcommand{\BEA}{\begin{eqnarray}} 
\newcommand{\EEA}{\end{eqnarray}} 
 
\renewcommand{\d}{{\rm d}} 
 
\newcommand{\ep}{\varepsilon}

\newcommand{\Wad}{\widetilde{W}}  
\newcommand{\aad}{\widetilde{a}}

\newcommand{\half}{\frac{1}{2}}

\newcommand{\ti}{t_{\rm i}} 
\newcommand{\tf}{t_{\rm f}} 
\newcommand{\Ri}{R_{\rm i}} 
\newcommand{\Rf}{R_{\rm f}} 
\newcommand{\Hi}{H_{\rm i}} 
\newcommand{\Hf}{H_{\rm f}}
\begin{document}  
\draft 
\title 
{Adiabatic processes need not correspond to optimal work}
\author{A.E. Allahverdyan$^{1,2)}$ 
and Th.M. Nieuwenhuizen$^{1)}$} 
\address{ 
$^{1)}$ Institute for Theoretical Physics, 
Valckenierstraat 65, 1018 XE Amsterdam, The Netherlands\\ 
$^{2)}$Yerevan Physics Institute, 
Alikhanian Brothers St. 2, Yerevan 375036, Armenia}  
 
\begin{abstract}
 
The minimum work principle states that work done on a thermally
isolated equilibrium system is minimal for the adiabatically slow
(reversible) realization of a given process.  This principle, one of
the formulations of the second law, is studied here for finite
(possibly large) quantum systems interacting with macroscopic sources
of work.  It is shown to be valid as long as the adiabatic energy
levels do not cross.  If level crossing does occur, counter examples
are discussed, showing that the minimum work principle can be violated
and that optimal processes are neither adiabatically slow nor
reversible.  
\end{abstract} 

 
\maketitle 
 
The second law of thermodynamics \cite{landau,balian,perrot},
formulated nearly one and half century ago, continues to be under
scrutinity \cite{QL2L,AN,NA}. While its status within equilibrium
thermodynamics and statistical physics is by now well-settled
\cite{landau,balian,perrot}, its fate in various border situations is
far from being clear. In the macroscopic realm the second law is a set
of equivalent statements concerning quantities such as entropy, heat,
work, etc.  However, in more general situations these statements need
not be equivalent and some, e.g. those involving entropy, may have
only a limited applicability\cite{AN,NA}.  In contrast to entropy, the
concept of work has a well-defined operational meaning for finite
systems interacting with macroscopic work sources \cite{balian}.  It
is, perhaps, not accidental that Thomson's formulation of the second
law \cite{landau,balian,perrot} | no work can be extracted from an
equilibrium system by means of a cyclis process | was proven
\cite{ANThomson,bassetttasaki} both in quantum and classical
situation.

Here we study the minimum work principle which extends the Thomson's
formulation to non-cyclic
processes~\cite{landau,balian,perrot,bassetttasaki}, and provides a
recipe for reducing energy costs.  After formulating the principle and
discussing it for macroscopic systems, we investigate it for finite
systems coupled to macroscopic sources of work.  Its domain of
validity there is found to be large but definitely limited.  These
limits are illustrated via counterexamples.
 
{\it The setup.}  Consider a quantum system S which is thermally
isolated \cite{landau,balian,perrot}: it moves according to its own
dynamics and interacts with an external macroscopic work source.
This interaction is realized via time-dependence of some parameters
$R(t)=\{R_1(t), R_2(t),...\}$ of the system's Hamiltonian
$H(t)=H\{R(t)\}$. They move along a certain trajectory $R(t)$ which at
some initial time $\ti$ starts from $\Ri=R(\ti)$, and ends at
$\Rf=R(\tf)$. The initial and final values of the Hamiltonian are
$\Hi=H\{\Ri\}$ and $\Hf=H\{\Rf\}$, respectively.  Initially S is
assumed to be in equilibrium at temperature $T=1/\beta\geq 0$, that
is, S is described by a gibbsian density operator:
\BEA 
\label{gibbs} 
\rho(\ti)=\exp(-\beta \Hi)/Z_{\rm i}, 
\qquad 
Z_{\rm i}={\rm tr}\,e^{-\beta \Hi}. 
\EEA 
As usual, this equilibrium state is prepared by a weak interaction
between S and a macroscopic thermal bath at temperature $T$
\cite{landau,balian,NA}, and then decoupling S from the bath in order
to achieve a thermally isolated process \cite{landau,balian,perrot}.

The Hamiltonian $H(t)$ generates a unitary evolution: 
\BEA 
i\hbar\frac{\d}{\d t}{\rho}(t)=[\,H(t),\rho(t)\,],\qquad 
\rho(t)=U(t)\,\rho(\ti)\,U^\dagger (t), 
\label{evolution} 
\EEA 
with time-ordered 
$U(t)=\overleftarrow{\exp}[-\frac{i}{\hbar}\int_{\ti}^{t}\d s\, H(s)]$.
The work $W$ done on S reads \cite{landau,balian,perrot} 
\BEA 
W=\int_{\ti}^{\tf}\d t\,{\rm tr}\,[ 
\rho(t)\dot{H}(t)]={\rm tr}[\Hf\rho(\tf)]- 
{\rm tr}[\Hi\rho(\ti) ], 
\label{work} 
\EEA 
where we performed partial integration and 
inserted (\ref{evolution}). This is the energy increase  
of S, which 
coincides with the energy decrease of the source. 

{\it The principle.}    
Let S start in the state (\ref{gibbs}), and  let
$R$ move between $\Ri$ and $\Rf$ along a trajectory $R(t)$.  
The work done on S is $W$. 
Consider the adiabatically slow realization of this process: $R$ 
proceeds between the same values $\Ri$ and $\Rf$ and along the 
same trajectory, but now with a homogeneously vanishing velocity, 
thereby taking a very long time $\tf-\ti$, at the cost of 
an amount work $\Wad$.
The minimum-work principle then asserts
\cite{landau,balian,perrot}  
\BEA 
\label{2L} 
W\geq \Wad . 
\EEA 
This is a statement on optimality: if work has to be extracted from S,
$W$ is negative, and to make it as negative as possible one proceeds
with very slow velocity.  If during some operation work has to be
added ($W>0$) to S, one wishes to minimize its amount, and
operates slowly.  For thermally isolated systems, adiabatically slow
processes are reversible. This is standard if S is macroscopic
\cite{landau,balian,perrot}, and below it is shown to hold
for a finite S as well, where the definition of reversibility
extends unambiguously (i.e., without invoking entropy) \cite{perrot}.
 
In macroscopic thermodynamics the minimum work principle is derived 
\cite{landau,perrot} from certain axioms which ensure that, 
within the domain of their applicability, this principle is equivalent 
to other formulations of the second law.  Derivations in the context 
of statistical thermodynamics are presented in 
\cite{jar,narnhofer,bassetttasaki}. We discuss one of them now.

{\it The minimal work principle for macroscopic systems} is proven in two 
steps: first one considers the relative entropy ${\rm 
tr}[\rho(\tf)\ln\rho(\tf)-\rho(\tf)\ln\rho_{\rm eq}(\Hf)]$ between the 
final state $\rho(\tf)$ given by (\ref{evolution}) and an equilibrium 
state $\rho_{\rm eq}(\Hf)=\exp(-\beta \Hf)/Z_{\rm f}$, $Z_{\rm f}={\rm 
tr}\,e^{-\beta \Hf}$, a state  corresponding to the final Hamiltonian 
$\Hf$ and the same temperature $T=1/\beta$.
As follows from (\ref{evolution}), ${\rm 
tr}[\rho(\tf)\ln\rho(\tf)]= {\rm tr}[\rho(\ti)\ln\rho(\ti)]$. This 
combined with (\ref{gibbs}, \ref{work}) and 
the non-negativity of relative entropy \cite{balian} yields: 
\BEA 
\label{workfree-energy} 
W\geq F(\Hf)-F(\Hi)\equiv T\ln {\rm tr}\,e^{-\beta \Hi}- 
T\ln {\rm tr}e^{-\beta \Hf},  
\EEA 
where $F(\Hi)$ and $F(\Hf)$ are the gibbsian free energies corresponding 
to $\rho(\ti)$ and $\rho_{\rm eq}(\Hf)$, respectively. 
 
There are several classes of macroscopic systems for which one can
show that the free energy difference in the RHS of
(\ref{workfree-energy}) indeed coincides with the adiabatic work
\cite{narnhofer,jar,NA}.

{\it Finite systems.} For an arbitrary $N$-level quantum system S,
Eq.~(\ref{workfree-energy}) does not have the needed physical meaning,
since in general $F(\Hf)-F(\Hi)$ does not coincide with the the
adiabatic work. It is known\cite{flu} that for finite systems the
final density matrix $\rho(\tf)$ given by (\ref{evolution}) need not
coincide with $\rho_{\rm eq}(\Hf)=\exp(-\beta \Hf)/Z_{\rm f}$. This
fact was recently applied for certain irreversible processes\cite{japan}.

Thus we need an independent  derivation of  (\ref{2L}). 
Let the spectral resolution of $H(t)$ and 
$\rho(\ti)$ be 
\BEA 
H(t)=\sum_{k=1}^N\ep_k(t)|k,t\rangle\langle k,t|,\quad 
\langle k,t|n,t\rangle =\delta_{kn},\\ 
\label{khorezm} 
\rho(\ti)=\sum_{k=1}^N p_k|k,\ti\rangle\langle k,\ti|,\quad 
p_k=\frac{e^{-\beta \ep_k(\ti)}}{\sum_n e^{-\beta \ep_n(\ti)}}. 
\EEA 
At $t=\ti$ we order the spectrum as 
\BEA 
\ep_1(\ti)\leq ...\leq \ep_N(\ti)\,\Longrightarrow\, 
p_1\geq...\geq p_N. 
\label{jalaledin} 
\EEA 
For $\ti\leq t\leq \tf$ we expand on the complete set $|n,t\rangle$: 
\BEA 
\label{chingiz} 
U(t)|k,\ti\rangle=\sum_{n=1}^N a_{kn}(t) \, 
e^{-\frac{i}{\hbar}\int_{\ti}^{t}\d t'\,\ep_n(t')}\,  
|n,t\rangle, 
\EEA 
and use (\ref{work}) to obtain: 
\BEA 
\label{suomi0} 
W=\sum_{k,n=1}^N 
|a_{kn}(\tf)|^2\,p_k\,\ep_n(\tf)-\sum_{k=1}^N p_k\,\ep_k(\ti). 
\EEA 
A similar formula can be derived to express the adiabatic work
$\Wad$ in coefficients  $\aad_{kn}(\tf)$.
From the definition $|a_{kn}(\tf)|^2=|\langle n,\tf|U|k,\ti\rangle|^2$ 
it follows that 
\BEA 
\label{bek} 
\sum_{k=1}^{N}|a_{kn}(\tf)|^2=\sum_{n=1}^{N}|a_{kn}(\tf)|^2=1. 
\EEA 
With help of the identity: 
$\sum_{n=1}^N \ep_n x_n=\ep_N\sum_{n=1}^Nx_n- 
\sum_{m=1}^{N-1}[\ep_{m+1}-\ep_{m}]\sum_{n=1}^m x_n$, we obtain
using (\ref{suomi0}, \ref{bek}) the general formula for the 
difference between non-adiabatic and adiabatic work:
\BEA 
\label{suomi1} 
W-\Wad=&&\sum_{m=1}^{N-1}[\ep_{m+1}(\tf)-\ep_{m}(\tf)]\,\Theta_m,\\ 
\Theta_m\equiv&&\sum_{n=1}^{m}\sum_{k=1}^N p_k 
(\,|\aad_{kn}(\tf)|^2-|a_{kn}(\tf)|^2). 
\label{suomi2} 
\EEA 

To understand the meaning of this formula, let us first
assume that the ordering (\ref{jalaledin}) is kept at $t=\tf$: 
\BEA 
\label{20'} 
\ep_1(\tf)\leq ...\leq \ep_N(\tf). 
\EEA 
If different energy levels did not cross each other
(and equal ones do not become different), Eq. (\ref{20'})
is implied by Eq. (\ref{jalaledin}).  According to non-crossing rule
\cite{mead1}, if $H\{R\}$ is real and only one of its parameters is
varied with time, (\ref{20'}) is satisfied for any discrete-level
quantum system: level-crossing, even if it happens in model-dependent
calculations or due to approximate symmetry, does not survive
arbitrary small perturbation where it is substituted by avoided
crossing (for a more general $H\{R\}$ the conditions prohibiting
level-crossing are more restrictive; see \cite{mead1}). 
No level-crossings and natural conditions of smoothmess of $H(t)$
are sufficient for the standard quantum adiabatic theorem 
\cite{standardad} to ensure
\BEA 
\label{abu} 
\aad_{kn}(\tf)=\delta_{kn}. 
\EEA 
Combined with (\ref{jalaledin}, \ref{abu}), Eq.(\ref{suomi2}) brings  
 \BEA 
\label{khan} 
\Theta_m=\sum_{k=1}^{m}p_k(1-\sum_{n=1}^m |a_{kn}(\tf)|^2)- 
\sum_{n=1}^{m}\sum_{k=m+1}^N p_k|a_{kn}(\tf)|^2\quad\nonumber\\ 
\geq p_m\left[m-\sum_{k=1}^{m}\sum_{n=1}^m |a_{kn}(\tf)|^2- 
\sum_{n=1}^{m}\sum_{k=m+1}^N |a_{kn}(\tf)|^2\right]=0.\quad\nonumber 
\EEA 
Eqs.~(\ref{suomi0}, \ref{suomi1}, \ref{20'}) together with
$\Theta_m\geq 0$ extend the minimum work principle (\ref{2L}) to cases
where the adiabatic work is not equal to the difference in free
energies.
 
{\it Level crossing.} The above non-crossing condition raises 
the question: Is the minimum work principle
also valid if the adiabatic energy levels cross?  
Before addressing this question in detail, let us mention some
popular misconceptions which surround the level-crossing problem: 1)
The no-crossing rule is said to exclude all crossings. This is
incorrect as the exclusion concerns situations where, in particular,
only one independent parameter of a real Hamiltonian $H\{R\}$ is
varied~\cite{mead1}. Two parameters can produce robust level-crossing
for such Hamiltonians.  2) It is believed that once levels can cross,
$\Delta\ep\to 0$, the very point of the adiabatic theorem disappears
as the internal characteristic time $\hbar/\Delta\ep$ of S is
infinite. This view misidentifies the proper internal time as seen
below; see also \cite{avron} 
in this context. 3) It is sometimes believed
that crossing is automatically followed by a population inversion. We
shall find no support for that.
 
As a first example we consider a spin-$1/2$ particle  
with Hamiltonian 
\BEA 
\label{sos} 
H(t)=h_1(s)\sigma_1-h_3(s)\sigma_3,\qquad 
s=t/\tau, 
\EEA 
where $\sigma_1$, $\sigma_3$ and $\sigma_2=i\sigma_1\sigma_3$ are
Pauli matrices, and where $s$ is the reduced time with $\tau$ being
the characteristic time-scale.  The magnetic fields $h_1$ and $h_3$
smoothly vary in time. Assume that {\it i)} for $s\to s_{\rm i}<0$ and
for $s\to s_{\rm f}>0$, $h_1(s)$ and $h_3(s)$ go to constant values
sufficiently fast; {\it ii)} for $s\to 0$ one has: $h_1(s)\simeq
\alpha_1 s^2$, $h_3(s)\simeq -\alpha_3 s$, where $\alpha_1$ and
$\alpha_3$ are positive constants.  {\it iii)} $h_1(s)$ and $h_3(s)$ are
non-zero for all $s$, $s_{\rm i}\leq s\leq s_{\rm f}$, except $s=0$.
Not all these points are needed, but we choose them for
clarity. Generalizations are indicated below.  One writes (\ref{sos})
as
\BEA H= 
\left(\begin{array}{rr} 
-h_3(s)&h_1(s) \\ 
h_1(s)&h_3(s) 
\end{array}\right)=\ep_1(s) 
\left(\begin{array}{rr} 
\cos\theta(s)&\sin\theta(s) \nonumber\\ 
\sin\theta(s)&-\cos\theta(s) 
\end{array}\right), 
\EEA 
where $\theta(s)=-\arctan[{h_1(s)}/{h_3(s)}]$  is a parameter  
in the interval $-\pi/2<\theta<\pi/2$, and where 
\BEA 
\ep_1(s)={\rm sg}(s)\sqrt{h^2_3(s)+h^2_1(s)},\quad 
\ep_2(s)=-\ep_1(s) 
\EEA 
are the adiabatic energy levels which cross at $s=\theta(s)=0$
($\sqrt{\dots}$ is defined to be always positive).
The adiabatic eigenvectors are, 
$H(s)|k,s\rangle=\ep_k(s)|k,s\rangle$, $k=1,2$, 
\BEA 
|1,s\rangle= 
\left(\begin{array}{r} 
\cos\frac{1}{2}{\theta(s)} \\ 
\\ 
\sin\frac{1}{2}{\theta(s)}  
\end{array}\right),\quad 
|2,s\rangle= 
\left(\begin{array}{r} 
-\sin\frac{1}{2}{\theta(s)} \\ 
\\ 
\cos\frac{1}{2}{\theta(s)}  
\end{array}\right). 
\label{sosend} 
\EEA 
Both the eigenvalues and the
eigenvectors are smooth functions of $s$. 
Eq.~(\ref{20'}) is not valid, and  
(\ref{suomi0}--\ref{suomi2}) imply:
\BEA \label{WminWadcross}
W-\Wad=-2\sqrt{h^2_1(s_{\rm f})+h^2_3(s_{\rm f})}\,\Theta_1, 
\quad \tau\,s_{\rm f}=\tf. 
\EEA 
Naively this already proves the violation. More carefully,
our strategy is now to confirm (\ref{abu}) in the adiabatic limit  
$\tau\to \infty$ and thus to confirm that $\Theta_1>0$,
implying that the minimum work principle is indeed violated.  
To this end we apply the standard adiabatic perturbation 
theory \cite{standardad}. Substituting (\ref{chingiz}) into  
(\ref{evolution}) one has: 
\BEA 
\label{shaman} 
\dot{a}_{kn}=-\sum_{m=1}^N a_{km}(t)\,
e^{\frac{i}{\hbar}\int_{\ti}^t\d t'[\ep_{n}(t')-\ep_{m}(t')]
}\, 
\langle n|\partial_t|m\rangle.
\EEA 
As $|1\rangle$ and $|2\rangle$ in (\ref{sosend}) are real, 
$\langle n|n\rangle=1$ implies $\langle n|\partial_t|n\rangle=0$. 
Since $\langle n|\partial_t|m\rangle 
=\frac{1}{\tau}\langle n|\partial_s|m\rangle$ the RHS of (\ref{shaman}) 
contains a small parameter ${1}/{\tau}$. 
It is more convenient to introduce new variables: 
$a_{kn}(t)=\delta_{kn}+b_{kn}(t)$, $b_{kn}(\ti)=0$. 
To leading order in $1/\tau$,
$b_{kn}$ can be neglected in the RHS of (\ref{shaman}), yielding for
$a_{k\not =n}(\tf)=b_{k\not =n}(\tf)$:
\BEA 
|a_{k\not =n}(\tf)|^2
=\left|\int_{s_{\rm i}}^{s_{\rm f}}\d s\,e^{ 
\frac{i\tau}{\hbar}\int^s_{s_{\rm i}}\d u[\ep_k(u)-\ep_n(u)]} 
\langle n|\partial_s|k\rangle\right|^2,~ 
\label{w3} 
\EEA 
while $|a_{kk}(\tf)|^2=1-\sum_{n\not =k}|a_{kn}(\tf)|^2$. 
In (\ref{w3}) we put 
$s\tau=t$, $s'\tau=t'$. For our model 
(\ref{sos}--\ref{sosend}), $\int_{s_{\rm i}}^s\d u[\ep_1(u)-\ep_2(u)]= 
2\int_{s_{\rm i}}^s\d u\,\ep_1(u)$ has only one extremal point, at $s=0$.  
We also have from (
\ref{sosend}) 
\BEA 
\langle 2|\partial_s|1\rangle= 
\frac{\theta'}{2}=\half\,\frac{h_1h'_3-h_3h'_1}{h^2_3+h^2_1},\quad 
\theta'\equiv\frac{\d \theta}{\d s}. 
\EEA 
For large $\tau$ the integral in 
(\ref{w3}) can be calculated with use of the saddle-point method: 
\BEA 
\label{w7} 
|a_{12}(\tf)|^2 
=\frac{\pi\hbar}{\tau}\left.
\left[
\frac{\langle 2|\partial_s|1\rangle^2
\sqrt{h^2_1+h^2_3}}{h_1h'_1+h_3h'_3}\right]\right|_{s=0}
=\frac{\pi\hbar\alpha_1^2}{4\tau\alpha_3^3}.~~
\EEA 
Eqs.~(\ref{w3}, \ref{w7}) extend the adiabatic theorem (\ref{abu}) for
the level-crossing situation. More general statements of similar adiabatic
theorems can be found in Ref.~\cite{avron}.  Inserting
$\Theta_1=(p_1-p_2)|a_{12}(\tf)|^2>0$ in Eq.~(\ref{WminWadcross})
confirms the violation of the minimum work principle. Eq.~(\ref{w7})
shows that the role of the proper internal characteristic time is
played by $\hbar \alpha_1^2/\alpha_3^3$ rather than by
$\hbar/(\ep_1-\ep_2)$.
 
More generally, if $\sqrt{h_3^2(s)+h_1^2(s)}$ is a smooth function for
all real $s$ (e.g., it is not $\propto |s|$), there are no crossings of
eigenvalues and (\ref{2L}) is valid.  If both $h_1(s)$ and $h_3(s)$
are linear for $s\to 0$, the leading term presented in (\ref{w7})
vanishes due to $\langle 2|\partial_s|1\rangle^2|_{s=0}=0$, and one
needs the second-order in the saddle-point expansion, to be compared
with the second-order term of the adiabatic perturbation theory.  This
leads to the same physical conclusions as (\ref{w7}) did, but with
$|a_{12}(\tf)|^2\sim \tau^{-3}$.

One can calculate $|a_{kn}(\tf)|$ yet in another limiting case, where
the characteristic time $\tau$ is very short.  It is well-known
\cite{standardad} that in this limit energy changes can be
calculated with frozen initial state of S. For the present situation
this leads from (\ref{sosend}) to
$|a_{12}(\tf)|^2=|a_{21}(\tf)|^2=|\langle 1,t_{\rm f}\,|\,2,t_{\rm
i}\rangle|^2 =\sin^2\half[\theta(\tf)-\theta(\ti)]$, and thus to
$\Theta_1=(p_1-p_2)\sin^2\half[\theta(\tf)-\theta(\ti)]$, again
positive.

\begin{figure}[bhb] 
\vspace{-0.3cm} 
\includegraphics[width=5cm]{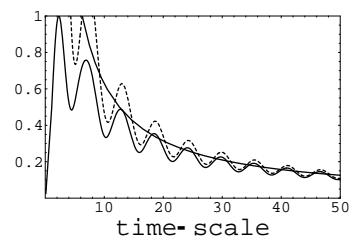} 
\vspace{-0.3cm} 
\caption{Amplitude $|a_{12}(s_{\rm f})|^2$ versus the time-scale 
$\tau$ for $s_{\rm i}=-1.5$, $s_{\rm f}=1.5$ and $\omega=1$. 
Full oscillating curve: the exact value which can reach unity.
Dotted curve: result from a first-order adiabatic perturbation 
theory. The smooth curve presents a saddle-point approximation 
analogous to (\ref{w7}).} 
\label{bobo} 
\end{figure}

{\it Exactly solvable model with level crossing}. 
Consider a two-level system with Hamiltonian 
\BEA 
\label{navu} 
H(t)=\hbar\omega 
\left(\begin{array}{rr} 
s\cos^2 s & ~~\half s\sin 2s \\ 
\\ 
\half s\sin 2s & ~~s\sin ^2s 
\end{array}\right), \qquad s=\frac{t}{\tau}, 
\EEA 
where $\tau$ is the characteristic time-scale, and $\omega$ is a constant. 
For $s_{\rm i}<0$ denote the adiabatic energy 
levels as $\ep_1(s_{\rm i})=\hbar\omega s_{\rm i}<\ep_2(s_{\rm i})=0$. 
They cross when $s$ passes through zero. 
Eq.~(\ref{shaman}) for this model can be solved exactly in terms 
of hypergeometric functions. Postponing the detailed discussion,  
we present in Fig. \ref{bobo}  
the behavior of $|a_{12}(s_{\rm f})|^2$ as a function 
of $\tau$. 
Since from (\ref{suomi1}, \ref{navu}) one has for the present model 
$W-\Wad=-\hbar\omega s_{\rm f}\,|a_{12}(s_{\rm f})|^2\, 
\tanh (\half\beta\hbar\omega\, |s_{\rm i}|\,)$, 
violations of the minimum work principle exist 
for $s_{\rm f}>0$, and they are maximal for $|a_{12}(s_{\rm 
f})|^2\to 1$.  This is seen to be the case in Fig. \ref{bobo} for some 
$\tau$ near $\tau=1$.  Notice that both the first-order perturbation 
theory and the saddle-point approximation are adequately reproducing 
$|a_{12}(s_{\rm f})|^2$ for $\tau \gtrsim 10$. 
 
Let S has a finite amount of levels, and two of them cross.  For
quasi-adiabatic processes ($\tau$ is large but finite) the transition
probability between non-crossing levels is exponentially small
\cite{standardad,pechu}, while as we saw it has power-law smallness
for two crossing levels.  Then one neglects in (\ref{suomi0}) the
factors $|a_{k\not =n}(\tf)|^2$ coming from any non-crossed levels $k$
and $n$, and the problem is reduced to the two-level situation. Thus
already one crossing suffices to detect limits of the minimum work
principle.  The reduction to the two-level situation takes place also in a
macroscopic system which has few discrete levels at the bottom of a
continuous spectrum, since for low temperatures these levels can
decouple from the rest of the spectrum.

{\it Cyclic processes and reversibility.}  The above results do not 
imply any violation of the second law in Thomson's 
formulation\cite{ANThomson}: no work is extracted from S during a cyclic 
process, $W_{\rm c}\geq 0$. 
We illustrate its general proof in the context of the level crossing
model given by (\ref{sos}--\ref{sosend}).  Assume that the trajectory 
$R(t)=(h_1(t),h_2(t)\,)$ described there is supplemented by another 
trajectory $R'(t)$ which brings the parameters back to their initial 
values $(h_1(\ti), h_3(\ti)\,)$ so that the overall process $R+R'$ is 
cyclic. If $R'$ crosses the levels backwards, then at the final 
time of $R'$ Eq.~(\ref{20'}) is valid, and 
(\ref{suomi0}--\ref{abu}) imply:
\BEA
W_{\rm c}=|a_{12}|^2(p_1-p_2)[\ep_2(\ti)-\ep_1(\ti)]
\geq \Wad_{\rm c}=0,
\label{trombon}
\EEA
where $|a_{12}|^2\leq 1$ now corresponds to the full cyclic process
$R+R'$.  Eq.~(\ref{trombon}) confirms the intuitive expectation that
non-adiabatic process are less optimal.  In particular, this is valid
if $R'$ is exactly the same process $R$ moved backwards with the same
speed.  Then
$\Wad_{\rm c}=0$ means that $R$ is a reversible process in the
standard thermodynamical sense \cite{landau,balian,perrot}.  If $R'$
does not induce another level crossing, i.e., $h_1(s)$ and $h_2(s)$ 
in Eq.~(\ref{sos}) return
to their initial values without simultaneously crossing zero, then
$\ep_1(\tf)=\ep_2(\ti)$, $\ep_2(\tf)=\ep_1(\ti)$ and
Eqs. (\ref{suomi0}, \ref{abu}) imply
\BEA
\Wad_{\rm c}=(p_1-p_2)[\ep_2(\ti)-\ep_1(\ti)]
\geq W_{\rm c}=|a_{11}|^2\,\Wad_{\rm c}>0.\nonumber
\EEA
In contrast to (\ref{trombon}), non-adiabatic processes are more
optimal if $R+R'$ contains one level-crossing (or an odd number of them). 
We thus have found here a violation of the minimum work principle
for a cyclic process.
\\ \indent 
{\it In conclusion}, we have studied the minimum work principle for
finite systems coupled to external sources of work.  As compared to
other formulations of the second law, this principle has a direct
practical meaning as it provides a recipe for reducing energy costs of
various processes. We gave its general proof and have shown that it
may become limited if there are crossings of adiabatic energy levels:
optimal processes need to be neither slow nor reversible. 
Already one crossing suffices to note violations of the principle.
If this is the case, the optimal process occurs at some finite speed.
\\ \indent 
Level-crossing was observed, e.g., in molecular and chemical
physics~\cite{chem}.  It is not a rare effect\cite{rare}: if the
number of externally varied parameters 
is larger then two, then for typical spectra level
crossings are even much more frequent than avoided crossings\cite{rare}.
It is possible that 
the presented limits of the minimum work principe may serve as a test
for level crossings.
\\ \indent 
Together with the universal validity of Thomson's formulation of the
second law \cite{ANThomson,bassetttasaki}, 
the limits of the principle imply that
the very equivalence between various formulations of the second law
may be broken for a finite system coupled to macroscopic sources of
work: different formulations are based on different physical
mechanisms and have different ranges of validity.  Similar results on
non-equivalence of various formulations of the second law were found
in Ref.~\cite{AN,NA}, where for a quantum particle coupled to a macroscopic
thermal bath, it was shown that some formulations,
e.g., Clausius inequality and positivity of the energy dispersion rate, 
are satisfied at sufficiently high temperatures of the bath, but
can be invalid at low temperatures, that is, in the quantum regime.
\\ \indent 
The work is supported by FOM/NWO (Netherlands).


\vspace{-0.7cm} 
\begin{thebibliography}{99} 
 
\bibitem{landau}L.D. Landau and E.M. Lifshitz, {\it Statistical 
Physics, I}, (Pergamon Press Oxford, 1978). 
 
\bibitem{balian} 
R. Balian, {\it From Microphysics to Macrophysics},  
volume I, (Springer, 1992). 
 
\bibitem{perrot}P. Perrot, {\it A to Z of Thermodynamics}, (Oxford 
University Press, 1998). 
 
\bibitem{QL2L} {\it Quantum Limits to the Second Law}, ed. D.P. Sheehan, 
(AIP Conf. Proc. 643, 2002) 
 
\bibitem{AN} A.E. Allahverdyan and Th.M. Nieuwenhuizen, Phys. 
Rev. Lett {\bf 85}, 1799 (2000); J. Phys. A, {\bf 36}, 875 (2003); 
Phys. Rev. E {\bf 64} 056117 (2001). 
 
\bibitem{NA} Th.M. Nieuwenhuizen and A.E. Allahverdyan, 
Phys. Rev. E {\bf 66}, 036102 (2002) 

\bibitem{ANThomson} 
G.N. Bochkov and Yu.E. Kuzovlev, Sov. Phys. JETP, {\bf 45}, 125 (1977).
W. Pusz and S.L. Woronowicz, Comm. Math. Phys., {\bf 58}, 273 (1978).
A. Lenard, J. Stat. Phys., {\bf 19}, 575 (1978).
J. Kurchan, cond-mat/0007360. 
A.E. Allahverdyan and Th.M. Nieuwenhuizen, Physica A {\bf 305}, 542
(2002).
 
\bibitem{bassetttasaki}I.M. Bassett, Phys. Rev. A {\bf 18}, 2356 (1978).

H. Tasaki, cond-mat/0009206. 

\bibitem{narnhofer} 
H. Narnhofer and W. Thirring, Phys. Rev. A, {\bf 26}, 3646 (1982). 



\bibitem{jar} 
K. Sekimoto and S. Sasa, J. Phys. Soc. Jpn. {\bf 66}, 3326 (1997).
C. Jarzynski, Phys. Rev. Lett. {\bf 78}, 2690 (1997).
G.E. Crooks, Phys. Rev. E {\bf 60}, 2721 (1999). 
C. Maes, J. Stat. Phys. {\bf 95}, 367 (1999). 
H. Tasaki, cond-mat/0009244. 

\bibitem{flu}
L. J. Broer, Physica {\bf 17}, 531 (1951). 
R. D. Mountain, Physica {\bf 30}, 808 (1964). 
M.J. Klein, Phys. Rev. {\bf 86}, 807 (1952).
R.M. Wilcox, Phys. Rev. {\bf 174}, 624 (1968).

\bibitem{japan}K. Sato, K. Sekimoto, T. Hondou, and F. Takagi,
Phys. Rev. E {\bf 66}, 016119 (2002).
H. Tasaki, cond-mat/0008420.


 


\bibitem{mead1} H.C. Longuet-Higgins, Proc. R. Soc. A, {\bf 344}, 147 
(1975). C.A. Mead, J. Chem. Phys., {\bf 70}, 2276 (1979).  

 
\bibitem{standardad}A. Galindo and P. Pascual, {\it Quantum Mechanics}, 
(Springer-Verlag, 1991).  


\bibitem{avron} M. Born and V. Fock, Z. Phys, {\bf 51}, 165 (1928).
G.A. Hagedorn, Ann. Phys., {\bf 196}, 278 (1989). 
J.E. Avron and A. Elgart, Comm. Math. Phys., 
{\bf 203}, 445 (1999).
 


 
\bibitem{pechu}J. Hwang and P. Pechukas, J. Chem. Phys., 
{\bf 67}, 4640 (1977). Y.N. Demkov, et al., 
Phys. Rev. A, {\bf 18}, 2089 (1978). 

\bibitem{rare}D.G. Truhlar and A.C. Mead, Phys. Rev. A {\bf 68},
032501 (2003).

\bibitem{chem}D.R. Yarkony, Rev. Mod. Phys., {\bf 68} (1996). 
H. Koppel, Chem. Phys. {\bf 77}, 359 (1983). 
H.D. Meyer, {\it ibid}. {\bf 82}, 199 (1983). 
C.A. Mead and D.G. Truhlar, J. Chem. Phys., 
{\bf 70}, 2284 (1979). 
 
 
 
 

\end{thebibliography}
\end{document}